\documentclass[aps,prb,reprint,amsmath,amssymb]{revtex4-2}
\usepackage{bm}
\usepackage{graphicx}

\begin{document}


\title{Energy Band Engineering of Periodic Scatterers by Quasi-1D Confinement}

\author{J. I. Kim}
\email{kim@unifesp.br}
\affiliation{Departamento de F\'\i sica,
Instituto de Ci\^encias Ambientais, Qu\'imicas e Farmac\^euticas \\ 
Universidade Federal de S\~ao Paulo, 
Rua S\~ao Nicolau 210, 09913-030, Diadema, SP, Brazil.}


\begin{abstract}
A mechanism to modify the energy band structure is proposed by considering a chain of periodic scatterers forming a 
linear lattice around which an external cylindrical trapping potential is applied along the chain axis. When this
trapping (confining) potential is tight enough, it may modify the bound and scattering states of the lattice  
potential, whose three-dimensional nature around each scattering center is fully taken into account and not
resorting to zero-range pseudo-potentials. Since these states contribute to the formation of the energy bands, such
bands could thereby be continuously tuned by manipulating the confinement without the need to change the lattice
potential. In particular, such dimensionality reduction by quantum confinement can close band gaps either at the
center or at the edge of the momentum $k$-space.  
\end{abstract}

\maketitle

\section{\label{sec:intro}Introduction}
Some physical properties of a material generally depend on its electronic band structure, particularly, whether it
behaves as a (band) insulator, a semiconductor or a metal~\cite{ashcroft1976}. In this regard, rarely can one and
the same material be freely driven to become, e.g., either an insulator in one situation or a metal in another, due
to the constraint imposed, for example, by the fixed crystal lattice, which determines much of the energy bands.
However, designing a material with tunable band structure but without changing its lattice, on the other hand,
would be highly desirable for both basic research and technological applications.  

Ultracold atoms and optical traps~\cite{grimm2000} have allowed the simulation of several physical
systems~\cite{bloch2008,giorgini2008,lewenstein2012,krutitsky2016}, including also many-body effects. The
single-particle band structure in an optical lattice can then be tuned by changing the laser parameters like its
intensity and wave length, or by changing the beams' configuration such as tilting the relative angle between the
directions of the counter-propagating beams~\cite{peil2003,hadzibabic2004,fallani2005} or setting a relative
non-zero angle between their polarization axes (the so called lin-$\theta$-lin
configuration~\cite{krutitsky2016,jessen1996,grynberg2001}). Although powerful and very convenient, such band
structure tuning techniques with optical lattices amount, however, to changing the lattice structure, i.e, to
changing the material itself such as its composition (the laser intensity or detuning driving the trap depth of
potential minima), size, periodicity, etc. 

Several studies show or predict that the band structure of many different materials or systems can in fact be
engineered by effectively changing the lattice structure in one way or another or by applying external fields. For
instance, one can use controlled impurity doping (e.g. in one-dimensional gold atomic wires~\cite{choi2008}),
chemical functionalization (e.g. hydrogenating graphene to obtain graphane~\cite{elias2009}), mechanical straining
(e.g. in two-dimensional graphene monoxide~\cite{pu2013}), mechanical deformation (e.g. radial deformation of
carbon nanotubes~\cite{gulseren2002}), electrically gating bilayer graphene~\cite{min2007,castro2007,zhang2009} 
or cutting nanoribbons into different widths~\cite{son2006,han2007,li2008}. 
This latter case is of particular interest here, as it involves geometric quantum confinement, which occurs also 
in some quantum dot lattices for mesoscopic electrons, in which external walls confine the (otherwise free)
internal motion of the particle within the dots and one could also observe related effects on the system energy
profile by subjecting them to external electromagnetic fields~\cite{munoz2005,drouvelis2007,morfonios2009}.

Geometric quantum confinement is also applied here by analysing a one-dimensional (1D) 
model~\cite{giamarchi2004,yurovsky2008} akin to ultracold atoms in a 1D optical lattice. Rather than manipulating
the lattice potential $V_{latt}$ itself, the band strucuture is then engineered with an external confining
potential $U$ with waveguide-like cylindrical symmetry around the 1D chain of scatterers of $V_{latt}$. By
quasi-1D is meant that one accounts locally for the physical 3D nature of both $U$ and $V_{latt}$ around each
lattice site as is described in Sec.~\ref{sec:model}. The main idea stems from previous seminal results in the
context of two-body cold atomic collisions in low 
dimensionality~\cite{olshanii1998,yurovsky2008,dunjko2011,granger2004} demonstrating, first in the
$s$-wave~\cite{olshanii1998} and then in the $p$-wave approximations~\cite{granger2004}, that the scattering
properties of a single scatterer can be strongly modified when it is placed in a tight atom 
waveguide. Here this low dimensional physics is generalized to a periodic chain of scatterers and to include
simultaneously both the $s$- and $p$-scattering waves, thus complementing Ref.~\cite{negretti2014} without, however,
using pure 1D zero-range pseudo-potentials and explicitly revealing how the 1D effective parameters depend on the
physical 3D ones, specially on the length scale of the confining potential $U$. For this purpose, it suffices to
apply the analytical techniques developed in Refs.~\cite{kim2005,kim2006,kim2007}, although more complete
treatments~\cite{giannakeas2012,hess2014,hess2015} could also be used as well. As a result, the band structure
substantially changes driven by the confinement, with some gaps vanishing if the so-called dual 
confinement-induced resonance (CIR) is reached~\cite{kim2006,kim2007}, whereby the effective quasi-1D scattering is
totally suppressed. 

After detailing the present model in Sec.~\ref{sec:model}, its solution around a single lattice site is discussed
in Sec.~\ref{sec:bandconf} and extended to the whole lattice in Sec.~\ref{sec:bands}, where the energy bands are
calculated as a function of the confining potential length scale. In Sec.~\ref{sec:discussion}, possible
experimental realizations are proposed. 

\section{\label{sec:model}A Quasi-1D Model System}
We assume an infinitely long linear lattice, with lattice constant $a$, and oriented along the $z$-axis. The
periodic lattice potential then satisfies $V_{latt}(\bm r + n_3a\bm e_3) = V_{latt}(\bm r)$ for 
$n_3 = 0, \pm 1, \pm 2,...$, where $\bm r = (x, y, z)$ is the position vector from the site placed at the origin
(zeroth site) and $\bm e_3$ is the unit vector along the $z$-axis. Usually $V_{latt} < 0$, i.e. attractive. In the
most symmetric configuration, the confining potential may be given by a cylindrically symmetric function $U(\rho)$,
where $\rho = (x^2 + y^2)^{1/2}$ is the distance from the lattice symmetry $z$-axis. In general, $U(\rho)$ is taken
to be zero at $\rho = 0$ and to grow positive as $\rho$ increases. The Hamiltonian $H$ for a spinless particle of
mass $m$ is then  
\begin{equation}
\label{eq:latthamiltonian}
H = -\frac{\hbar^2}{2m}\nabla^2 + U(\rho)  + V_{latt}(\bm r).
\end{equation}
Two symmetries can be identified here, an external one brought about by the confinement and the local one
associated to each lattice site. It is by exploring the interplay between them that a band structure mechanism is
proposed. 

Under a few proof-of-concept approximations, an analytical solution can be readily obtained. Although many
functions $U(\rho)$ and other geometries for the confinement are possible (e.g. $U_1(x) + U_2(y)$), we adopt here
the simplest variant by approximating $U(\rho)$ by a square-well type function, namely, $U(\rho) = 0$ for 
$\rho < R_U$ and infinite otherwise, $R_U$ being the range of $U(\rho)$. A second important assumption is to
approximate $V_{latt}(\bm r)$ around each lattice site by a spherically symmetric potential $V$, namely 
\begin{subequations}
\begin{equation}
\label{eq:approxV}
V_{latt}(\bm r)\approx V(r),
\end{equation}
for $\bm r$ around the zeroth site, where $r = |\bm r| = (x^2 + y^2 + z^2)^{1/2}$, and 
$V_{latt}(\bm r)\approx V(|\bm r - n_3a\bm e_3|)$ for $\bm r$ around the $n_3$-th site and so forth, thus implying 
$R_V \ll a$, where $R_V$ is the range of $V$. As for the range $R_U$, we assume 
\begin{equation}
\label{eq:approxRvRua}
R_V  \ll R_U \ll a, 
\end{equation}
\end{subequations}
i.e. the lattice interaction occurs predominantly at the center of the confinement, which in turn is taken to be very
tight relative to the lattice dimension. It should be noted that these restrictions are enough for the present purposes 
but can be lifted in more involved analytical and numerical calculations. 

The solution $\Psi(\bm r)$ to Eq.(\ref{eq:latthamiltonian}) should obey Bloch's condition, one form of which is
$\Psi(\bm r + n_3a\bm e_3) = e^{in_3ak}\Psi(\bm r)$, for some $k$. A more convenient form is to shift $\bm r + n_3a\bm e_3
\longrightarrow \bm r$ to get  
\begin{equation}
\label{eq:bloch}
\Psi(\bm r) = e^{in_3ak}\Psi(\bm r - n_3a\bm e_3),
\end{equation}
so that the solution $\Psi(\bm r)$ around any lattice site, say at $n_3a\bm e_3$ (i.e. $\bm r$ around $n_3a\bm e_3$),
can be expressed in terms of the solution $\Psi(\bm r - n_3a\bm e_3)$ around the origin (i.e. $\bm r - n_3a\bm e_3$
around $\bm 0$). Therefore, one can focus on this latter solution (Sec.~\ref{sec:bandconf}) and than expand it via
Bloch's condition above to cover the whole lattice (Sec.~\ref{sec:bands}). 

\section{\label{sec:bandconf}Local States under Confinement}
In order to find the solution $\Psi(\bm r)\equiv\Psi_0(\bm r)$ around the site at the origin, we note that from
Eq.(\ref{eq:approxV}) the Schr\"odinger equation for $\Psi_0$ follows from Eq.(\ref{eq:latthamiltonian}), namely
\begin{equation}
\label{eq:0hamiltonian}
\left[ -\frac{\hbar^2}{2m}\nabla^2 + U(\rho)  + V(r) \right]\Psi_0 = E\Psi_0, \hspace{1em} |z| < a/2,
\end{equation}
where $E$ is the total energy and where two conflicting symmetries (those of $U$ and $V$) can be clearly seen. As a 
first treatment to illustrate the principle, we assume $E\equiv\hbar^2K^2/2m > 0$ and deal with running waves bound
only by the lateral confinement ($E<0$ would correspond to the case of states deeply bound to the lattice sites,
although $U$ could raise $E$ towards higher energíes if it is tight enough to reach such deep states). It turns out
that precisely this Eq.(\ref{eq:0hamiltonian}) has already been studied in the context of cold atom scattering
under confinement (see e.g. Refs.~\cite{kim2005,kim2006,kim2007,melezhik2007} directly related to 
the present technique), in which case $V$ is the atom-atom interaction potential and $U$ is the confining laser
optical potential, so that much of the analysis for cold atoms remains valid. However, Bloch's condition introduces
important changes, such as a different boundary condition. For this reason and for the sake of completeness, we
repeat details of the analysis.

Indeed, let $\varphi_n$, for $n = 0, 1, 2...$, be the orthonormalized eigenstates of $U$ regular at $\rho=0$ with
eigenvalues $\hbar^2q_n^2/2m>0$ and satisfying  
\begin{subequations}
\begin{equation}
\label{eq:Uhamiltonian}
\left[ -\left(\frac{\partial^2}{\partial x^2} + \frac{\partial^2}{\partial y^2}\right) + u \right]\varphi_n =
q_n^2\varphi_n, 
\hspace{0.5em} n = 0, 1, 2... ,
\end{equation}
where $u \equiv 2m\,U(\rho)/\hbar^2$. Since $V(r)$ in Eq.(\ref{eq:0hamiltonian}) does not depend on the azimuthal
angle $\phi$ (that encircling the $z$-axis) either, one can take both $\varphi_n = \varphi_n(\rho)$ and $\Psi_0$ as
axially symmetric. For example, in the specific case of a square-well $U$, one obtains Bessel functions $J_m$
\begin{equation}
\label{eq:phin}
\varphi_n(\rho) = \frac{N_n}{\pi^{1/2}R_U}J_0(q_n\rho),
\end{equation}
\end{subequations}
where $N_n\equiv|J_1(r_{n+1})|^{-1}$ and $r_{n+1}$ is the $(n+1)$-th root of $J_0$, with $q_nR_U=r_{n+1}$. A good
approximation to $q_n$ is given by Eq.(13b) in~\cite{kim2005}, namely, $q_n\approx (n + 3/4)\pi/R_U$. Decomposing
then $\Psi_0$ on such type of basis $\{\varphi_n\}$, 
\begin{equation}
\label{eq:Psi0}
\Psi_0(\bm r) = \sum_{n = 0}^{\infty}\psi_n(z)\varphi_n(\rho),
\end{equation}
substituting back into Eq.(\ref{eq:0hamiltonian}) and using the orthonormality of $\{\varphi_n\}$ gives for each
$\psi_n(z)$ 
\begin{equation}
\label{eq:psin}
\left(\frac{d^2}{dz^2} + k_n^2\right)\psi_n = \int dxdy \,\varphi_n^\ast(\rho)v(r)\Psi_0(\bm r),
\end{equation}
where $k_n^2 \equiv K^2 - q_n^2$ and $v(r) \equiv 2m\,V(r)/\hbar^2$. A general solution $G_n(z,z')$ to the 1D Green's
function equation $G_n''(z,z') + k_ n^2G_n(z,z') = -\delta(z - z')$ is 
\begin{equation}
\label{eq:1Dgreen}
G_n(z,z') = -\xi_{n+}\frac{e^{ik_n|z-z'|}}{2ik_n} - \xi_{n-}\frac{e^{-ik_n|z-z'|}}{2i(-k_n)},
\end{equation}
where $\xi_{n+} + \xi_{n-} = 1$. For now, the inward scattering wave $\xi_{n-}$ is also kept (as if to account for an
incoming flux from the lateral sites), but later we will be able to discard it based on boundary conditions, notably for 
$a\longrightarrow\infty$. Then $\psi_n$ can be written as 
\begin{eqnarray}
\label{eq:psingreen}
\psi_n(z) & = & A_n e^{ik_nz} + B_n e^{-ik_nz} \nonumber \\
          &   & \mbox{}  - \int d^3\bm r' G_n(z,z')\varphi_n^\ast(\rho')v(r')\Psi_0(\bm r'),
\end{eqnarray} 
where the first two terms are homogeneous solutions to the left-hand side (lhs) of Eq.(\ref{eq:psin}), representing
forward and backward running waves, respectively. One can now write Eq.(\ref{eq:Psi0}) as
\begin{eqnarray}
\label{eq:Psi0n}
\Psi_0(\bm r) & = & \sum_{n=0}^\infty\left( A_ne^{ik_nz} + B_ne^{-ik_nz}\right)\varphi_n(\rho) \\
              &   & - \sum_{n=0}^\infty\int d^3\bm r'G_n(z,z')\varphi_n^\ast(\rho')v(r')\Psi_0(\bm r')\varphi_n(\rho). 
\nonumber
\end{eqnarray}
Suppose now for simplicity that only the ground state channel $n=0$ is open, that is,
\begin{equation}
\label{eq:singlechannel}
0 \leq q_0^2 \leq K^2 \leq q_1^2.
\end{equation}
For $n\geq 1$, $k_n$ is then imaginary, say $k_n = i(q_n^2 - K^2)^{1/2}$, and thus $\xi_{n-}$ must vanish (so that
$\xi_{n+}=1$), otherwise the second term on the right-hand side (rhs) of Eq.(\ref{eq:Psi0n}) for $z\sim a/2$ would be
exponentially large as $e^{+a/2R_U}$ (see Eq.(\ref{eq:approxRvRua})), since $k_n\sim q_n \sim 1/R_U$ or larger and the
solution would diverge in the limit $a\rightarrow\infty$ of a single site; besides, if Eq.(\ref{eq:Psi0n}) is to be
finite for both $z\sim -a/2$ and $z\sim +a/2$, the $A_n$ and $B_n$, respectively, must also vanish. We have thus
\begin{eqnarray}
\label{eq:Psi0+}
\Psi_0(\bm r) & = & \left( A_0e^{ik_0z} + B_0e^{-ik_0z}\right)\varphi_0(\rho) \\
              &   & - \int d^3\bm r' G_0(z,z')\varphi_0^\ast(\rho')v(r')\Psi_0(\bm r')\varphi_0(\rho) \nonumber\\
              &   & + \sum_{n=1}^\infty\int d^3\bm r' \frac{e^{ik_n|z-z'|}}{2ik_n}
                        \varphi_n^\ast(\rho')v(r')\Psi_0(\bm r')\varphi_n(\rho), \nonumber
\end{eqnarray}
where we take $k_0\equiv +\sqrt{K^2 - q_0^2}>0$.

In order to impose Bloch's condition Eq.(\ref{eq:bloch}) (see also Eqs.(\ref{eq:continuity-bloch-a}) and
(\ref{eq:continuity-bloch-b})), we need to know the behaviour of $\Psi_0$ near the zero-th site boundaries $|z|\sim  
a/2$. In this region, $z$ predominates over $z'$, since $z'$ is limited to the range $R_V\ll a$ in the integrands
above. Hence, for $z\sim a/2$, $z-z'$ is positive and thus $|z-z'|= z - z'$, whereas for $z\sim -a/2$, $z-z'$ is
negative and thus $|z-z'|=-z + z'$, in other words, $|z-z'|=|z|\mp z'$ if $z\sim \pm a/2$. One notes also that each term
of the series for $n\geq 1$ in Eq(\ref{eq:Psi0+}) becomes exponentially small for $|z|\sim a/2$, scaling as
$e^{-a/2R_U}$ or less. As part of the approximations used here, we neglect them. In this way, Eq.(\ref{eq:Psi0+})
becomes for $z\sim\pm a/2$ 
\begin{subequations}
\begin{eqnarray}
\label{eq:Psi0border}
\Psi_0(\bm r) & \approx &  \left( A_0e^{ik_0z} + B_0e^{-ik_0z}\right)\varphi_0(\rho) \\
              &   & \hspace{1em} + \left(\xi_{0+}f_{0+}^\pm e^{ik_0|z|} + \xi_{0-}f_{0-}^\pm
                e^{-ik_0|z|}\right)\varphi_0(\rho), \nonumber 
\end{eqnarray}
the scattering amplitudes $f_{0+}^\pm$ and $f_{0-}^\pm$ being defined as
\begin{eqnarray}
\label{eq:f0+-}
f_{0+}^\pm & \equiv & \frac{1}{2ik_0}\int d^3\bm r'\left[ e^{\pm ik_0z'}\varphi_0(\rho')\right]^\ast v(r')\Psi_0(\bm r'),\\
f_{0-}^\pm & \equiv & \frac{1}{2i(-k_0)}\int d^3\bm r'\left[ e^{\pm i(-k_0)z'}\varphi_0(\rho')\right]^\ast v(r')\Psi_0(\bm
r'),\nonumber
\end{eqnarray}
\end{subequations}
the upper $(+)$ sign refering to $z\sim +a/2$ and the upper $(-)$ sign refering to $z\sim -a/2$. 

These amplitudes $f_{0+}^\pm$ and $f_{0-}^\pm$ depend on the behaviour of $\Psi_0$ in the region $R_V$ close to the
origin, where the spherical symmetry of $V$ prevails instead of the cylindrical one close to the borders 
$|z|\sim a/2$. It is more convenient then to replace the cylindrical basis vector $e^{\pm ik_0z}\varphi_0(\rho)$ by
a spherical basis, namely by trying
\begin{subequations}
\begin{equation}
\label{eq:ephi-jp}
e^{ik_0z}\varphi_0(\rho) = \sum_{l=0}^\infty [i^l(2l+1)\alpha_{0l}]\,j_l(Kr)P_l(\cos{\theta}),
\end{equation}
for some constants $\alpha_{0l}$, where $\theta$ is the polar angle, $j_l$ and $P_l$ are spherical Bessel functions
and Legendre polynomials, respectively, and the coordinate transformation is $z = r\cos{\theta}$ and 
$\rho = r\sin{\theta}$. If $U$ in Eq.(\ref{eq:Uhamiltonian}) is of the square-well type, $\alpha_{0l}$ can be
calculated exactly (see Eqs.(13) and (17) in~\cite{kim2005} and Eq.(11.3.49) in~\cite{morse1981}, Vol.II, \S 11.3)  
\begin{equation}
\label{eq:alpha0l}
\alpha_{0l} = \frac{1}{\pi^{1/2}d_U}P_l(k_0/K),
\end{equation}
\end{subequations}
where $d_U \equiv R_U/N_0$. When inserting
Eq.(\ref{eq:ephi-jp}) into Eqs.(\ref{eq:f0+-}), we use $-z = r\cos{(\pi - \theta)}$ and the property $P_l(\cos({\pi -
  \theta))} = (-1)^lP_l(\cos{\theta})$; in addition, we separate the $l$-summation into even $l=0,2,4...$ and odd
$l=1,3,5...$ groups, so that Eq.(\ref{eq:f0+-}) becomes
\begin{subequations}
\begin{eqnarray}
\label{eq:f0+}
f_{0+}^\pm & = & \sum_{l=\mathrm{even}}^\infty\frac{4\pi(2l+1)\alpha_{0l}^\ast}{2ik_0}\,T_l \\
          &   & \hspace{5em} \pm \sum_{l=\mathrm{odd}}^\infty\frac{4\pi(2l+1)\alpha_{0l}^\ast}{2ik_0}\,T_l, \nonumber\\
\label{eq:f0-}
f_{0-}^\pm & = & -\sum_{l=\mathrm{even}}^\infty\frac{4\pi(2l+1)\alpha_{0l}^\ast}{2ik_0}\,T_l \\
          &   & \hspace{5em}  \pm \sum_{l=\mathrm{odd}}^\infty\frac{4\pi(2l+1)\alpha_{0l}^\ast}{2ik_0}\,T_l, \nonumber
\end{eqnarray}
\end{subequations}
where $T_l$ are scattering amplitudes in the spherical basis
\begin{equation}
\label{eq:Tl}
T_l \equiv \int\!\frac{d^3\bm r'}{i^l4\pi}\left[j_l(Kr')P_l(\cos{\theta'})\right]v(r')\Psi_0(\bm r').
\end{equation}

For these $T_l$, one can not use Eq.(\ref{eq:Psi0border}), but needs $\Psi_0$ within the range
$R_V$. Following~\cite{kim2005,kim2006,kim2007}, the idea is to transform the rhs of the full solution
Eq.(\ref{eq:Psi0+}) using spherical coordinates, which will naturally bring about the amplitudes $T_l$. For this
purpose, we note that $r'\sim R_V\ll R_U$ and set $r\ll R_U$ as well, such that Eq.(\ref{eq:Psi0+}) becomes
\begin{equation}
\label{eq:Psi0+Gc}
\Psi_0(\bm r)  =  \Psi_{0i}(\bm r) - \int d^3\bm r'G_c(\bm r, \bm r')v(r')\Psi_0(\bm r'),
\end{equation}
where $\Psi_{0i}(\bm r)\equiv \left( A_0e^{ik_0z} + B_0e^{-ik_0z}\right)\varphi_0(\rho)$ and the axially symmetric Green's
function $G_c$ is, for $r,r'\ll R_U$,
\begin{eqnarray}
\label{eq:Gc}
G_c(\bm r,\bm r') & = & i(\xi_{0+} -
\xi_{0-})\varphi_0^\ast(\rho')\varphi_0(\rho)\frac{\cos{(k_0|z-z'|)}}{2k_0}\nonumber\\
 &  & \mbox{} - \varphi_0^\ast(\rho')\varphi_0(\rho)\frac{\sin{(k_0|z-z'|)}}{2k_0} \\
 &  & + \int_{q_1}^\infty\frac{qdq}{4\pi}J_0(q\rho')J_0(q\rho)\frac{e^{-\sqrt{q^2-K^2}|z-z'|}}{\sqrt{q^2-K^2}},\nonumber
\end{eqnarray}
where we used $k_n=+i(q_n^2 - K^2)^{1/2}$ for $n\geq 1$ and the third term, originally a discret summation over $n$, has
been replaced by its continuum limit (see Eq.(13b) et seq. in~\cite{kim2005}) valid for $r,r'\ll R_U$. Note that
Eq.(\ref{eq:Psi0+Gc}) can also be expressed alternatively by formally substituting $G_c$ by a non-axially symmetric
Green's function $G_u$ satisfying $[\nabla^2 - u(\rho) + K^2]G_u(\bm r, \bm r') = -\delta(\bm r - \bm r')$, such
that $2\pi G_c = \int d\phi'G_u$ since $G_c$, $v$ and $\Psi_0$ do not depend on $\phi'$ in Eq.(\ref{eq:Psi0+Gc}),
where the $\phi'$ integration can be made to yield the factor $2\pi$. As is discussed in
Sec.(IV.B) of~\cite{kim2005}, for $r,r'\ll R_U$, such that $u(\rho)\approx 0$, $G_u$ should differ from the
free-space 3D Green's function $G(\bm r, \bm r')$ satisfying 
$[\nabla^2 + K^2]G(\bm r,\bm r') = - \delta(\bm r - \bm r')$ by at 
most a homogeneous term $\Delta_u(\bm r, \bm r')$ satisfying $[\nabla^2 + K^2]\Delta_u(\bm r,\bm r') = 0$, so that
\begin{equation}
\label{eq:GcG}
G_c(\bm r,\bm r')\approx \int_0^{2\pi}\frac{d\phi'}{2\pi}G(\bm r,\bm r') + \Delta_c(\bm r,\bm r'),
\end{equation}
with $\Delta_c(\bm r,\bm r') \equiv \int d\phi'\Delta_u(\bm r, \bm r')/2\pi$ and where 
\begin{equation}
\label{eq:G}
G(\bm r,\bm r')\equiv \gamma_+\frac{e^{iK|\bm r - \bm r'|}}{4\pi|\bm r - \bm r'|}
                       + \gamma_-\frac{e^{-iK|\bm r - \bm r'|}}{4\pi|\bm r - \bm r'|}
\end{equation}
with $\gamma_+ + \gamma_- = 1$. In order to identify $\Delta_c(\bm r,\bm r')$ and these $\gamma$'s, one expands
$G(\bm r,\bm r')$ in cylindrical coordinates using (see, e.g.~\cite{morse1981}, Vol.I, Chap. 7, problem 7.9) 
\begin{eqnarray}
\label{eq:GJe}
\frac{e^{iK|\bm r - \bm r'|}}{4\pi|\bm r - \bm r'|} & = & -\sum_{m=0}^\infty(2 - \delta_{0,m})\cos{[m(\phi-\phi')]}\\
 &  & \times\int_0^\infty\frac{qdq}{4\pi}J_m(q\rho)J_m(q\rho')\frac{e^{i\sqrt{K^2-q^2}|z-z'|}}{i\sqrt{K^2-q^2}}\nonumber
\end{eqnarray}
with the correct branch $0\leq \mathrm{arg}\sqrt{K^2-q^2}<\pi$ and taking the complex conjugate to generate the
expansion of $e^{-iK|\bm r - \bm r'|}/4\pi|\bm r - \bm r'|$. We next substitute these expansions into the rhs of
Eq.(\ref{eq:GcG}), whose lhs in turn follows from Eq.(\ref{eq:Gc}), and compare both sides to get for $r,r'\ll R_U$
\begin{subequations}
\begin{eqnarray}
\label{eq:gammas}
\gamma_+ & = & \gamma_- = 1/2, \\
\label{eq:deltac}
\Delta_c(\bm r, \bm r') & \approx & -\int_0^{p_c}\frac{dp}{4\pi}J_0(q\rho')J_0(q\rho)e^{-p|z-z'|} \\
       &  & \mbox{} + i(\xi_{0+}-\xi_{0-})\varphi_0^\ast(\rho')\varphi_0(\rho)\frac{\cos{(k_0|z-z'|)}}{2k_0},\nonumber
\end{eqnarray}
\end{subequations}
where $q = \sqrt{K^2+p^2}$ and $p_c\equiv\sqrt{q_1^2-K^2}$. Here, Eq.(\ref{eq:gammas}) and Eq.(\ref{eq:deltac}) are
improvements to Eq.(16a) and Eq.(16b), respectively, of~\cite{kim2005} and follows the discussion given
in~\cite{kim2006,kim2007}. Physically, Eq.(\ref{eq:gammas}) accounts for an inward particle flux arising from
reflections of the outward scattered wave against the boundaries of the confinement. The 
integral in $\Delta_c$ stems from the lower limit $q_1$ in the $q$-integration in Eq.(\ref{eq:Gc}) and the limit $K$
implicit in the $q$-integration in Eq.(\ref{eq:GJe}), whereas a term involving $\sin{(\sqrt{K^2-q^2}|z-z'|)}$ and one
involving $\sin{(k_0|z-z'|)}$ have been neglected, since they are a factor $|z-z'|/R_U\ll 1$ smaller than the first and
second terms, respectively, of Eq.(\ref{eq:deltac}). Because of the modulus $|z - z'|$ rather than 
$z - z'$, $\Delta_c$  in Eq.(\ref{eq:deltac}) is not precisely an axially symmetric plane wave (in cylindrical
coordinates, with $\varphi_0$ being Bessel functions) satisfying the original requirement 
$[\nabla^2 + K^2]\Delta_c(\bm r,\bm r') = 0$, stemming from $\Delta_u$. On the other hand, this modulus plays an
important part in causing the decoupling of partial waves $l$ and $s$ for which $l + s$ is odd, as is discussed
when deriving Eq.(\ref{eq:Psi0jn}) below. In any case, detailed numerical calculations~\cite{kim2006,melezhik2007}
showed satisfactory agreements with such analytical approximations made here. For this reason, this $\Delta_c$ will
be kept in the following. 

The next step to calculate $T_l$ is to replace $G_c(\bm r,\rm r')$ in Eq.(\ref{eq:Psi0+Gc}) by the rhs of
Eq.(\ref{eq:GcG}), using the results Eqs.(\ref{eq:G}), (\ref{eq:gammas}) and (\ref{eq:deltac}). In this way, one can
safely expand the rhs of Eq.(\ref{eq:Psi0+Gc}) in spherical coordinates. For $\Psi_{0i}$ one uses directly
Eq.(\ref{eq:ephi-jp}). For $G$, one needs (see e.g. Eq.(11.3.44) in~\cite{morse1981}, Vol.II, \S 11.3)
\begin{eqnarray}
\label{eq:GjP}
\lefteqn{\frac{e^{iK|\bm r - \bm r'|}}{4\pi|\bm r - \bm r'|}  = \hspace{10em} r'<r} \\
             & & \frac{iK}{4\pi}\sum_{l=0}^\infty(2l+1)
                     \times \sum_{m=0}^l\epsilon_m\frac{(l-m)!}{(l+m)!}\cos{[m(\phi-\phi')]}\nonumber\\
             & & \times P_l^m(\cos{\theta'})P_l^m(\cos{\theta})j_l(Kr')[j_l(Kr) + i\,n_l(Kr)],\nonumber
\end{eqnarray}
where $\epsilon_m=1$ for $m=0$ and $\epsilon_m=2$ otherwise, $n_l$ is the spherical Neumann function and $P_l^m$ is
the associated Legendre 
function. As for $\Delta_c$ in Eq.(\ref{eq:deltac}), one rewrites $|z-z'|=z\sigma_{zz'}-z'\sigma_{zz'}$,
with $\sigma_{zz'}\equiv \mathrm{sign}(z-z')$; for its second term on the rhs one uses Eq.(\ref{eq:ephi-jp}) twice
(for $\bm r$ and then for $\bm r'$) and for its first term, on the other hand, one continues Eq.(\ref{eq:ephi-jp})
analytically to the imaginary $k_0\rightarrow ip$ axis to get (assuming square-well type $U$) 
\begin{equation}
\label{eq:eJ-jp}
e^{-pz}J_0(q\rho) = \sum_{l=0}^\infty i^l(2l+1)P_l(ip/K)j_l(Kr)P_l(\cos{\theta}),
\end{equation}
which is then also used twice. We now substitute these results into the rhs of Eq.(\ref{eq:GcG}) and cast
Eq.(\ref{eq:Psi0+Gc}) as an expansion in the 
spherical basis $\{j_lP_l\}$ valid for $r\ll R_U$. In doing so, one must carefully track signs such as $(-)^l$,
$(-)^s$, $(\sigma_{zz'})^l$ and $(\sigma_{zz'})^s$ and eliminate spurious couplings between even and odd angular
momenta arising from Eq.(\ref{eq:deltac}) (see discussion after Eq.(20) in~\cite{kim2005}), since 
$\langle l|U|s\rangle = 0$ if $l+s= \mathrm{odd}$, which then sets $(-)^{l + s} = (\sigma_{zz'})^{l+s} = +1$. As a
result, Eq.(\ref{eq:Psi0+Gc}) becomes for $r\ll R_U$ 
\begin{eqnarray}
\label{eq:Psi0jn}
\Psi_0(\bm r) & \approx & \sum_{l=0}^\infty i^l(2l+1)\left[\alpha_l + \gamma_l^{(1)} + i\gamma_l^{(2)}\right]
                                                                           j_l(Kr)P_l(\cos{\theta})\nonumber\\
              &  & + \sum_{l=0}^\infty i^l(2l+1)\left[KT_l\right]n_l(Kr)P_l(\cos{\theta}),
\end{eqnarray}
where $\alpha_l\equiv [A_0 + (-1)^lB_0]\,\alpha_{0l}$ and the $\gamma_l^{(m)}$'s are defined by 
$\gamma_l^{(m)}\equiv \sum_{s[l]}(2s+1)P_{ls}^{(m)}T_{s}$, with $s[l]$ being a sum over all even (odd) $s$ for a
given even (odd) $l$, and   
\begin{subequations}
\begin{eqnarray}
\label{eq:Pls1}
P_{ls}^{(1)} & \equiv & K\int_0^{p_c/K}dx\,P_l(ix)P_s(ix), \\
P_{ls}^{(2)} & \equiv & -(\xi_{0+}-\xi_{0-})\frac{2\pi}{k_0}\alpha_{0l}\alpha_{0s}.
\end{eqnarray}
\end{subequations}
We now set $R_V\ll r \ll R_U$ and recall that then the above $\Psi_0$ in Eq.(\ref{eq:Psi0jn}) should correspond to
a standard well-known spherical scattering solution emanating from $V$ (supposed of short range) and thus having
the form (see e.g. \cite{landau1980}, \S 132)
\begin{equation}
\label{eq:Psi0deltal}
\Psi_0(\bm r) \approx \sum_{l=0}^\infty c_l\left[ \cos{\delta_l}j_l(Kr) -
                                       \sin{\delta_l}n_l(Kr)\right]P_l(\cos{\theta})
\end{equation}
where $\delta_l$ is the physical 3D $l$-th angular momentum component scattering phase-shift. Comparing with
Eq.(\ref{eq:Psi0jn}) term by term and then eliminating the $c_l$'s, one gets for $T_l$ the following matrix
equation for $l=0,1,2...$ 
\begin{equation}
\label{eq:Tlmatrix}
-\left(K\cot{\delta_l}\right)\,T_l = \alpha_l + \sum_{s[l]}(2s+1)\left[P_{ls}^{(1)} + iP_{ls}^{(2)}\right]T_s
\end{equation}
which allows us to obtain $T_l$ and $f_{0\pm}^\pm$ in Eqs.(\ref{eq:f0+}) and (\ref{eq:f0-}), thus completing the
calculation of Eq.(\ref{eq:Psi0border}). Although higher partial waves could be collected from
Eq.(\ref{eq:Tlmatrix}), in the low energy condition $R_V\ll R_U$, it is usually a good approximation to retain only
the leading phase-shifts $\delta_0$ and $\delta_1$ (see also discussion in Sec.V.C of~\cite{kim2005}). Solving then
Eq.(\ref{eq:Tlmatrix}) for $T_0$ and $T_1$, one gets 
\begin{subequations}
\begin{eqnarray}
\label{eq:f0+T}
f_{0+}^\pm & = & -\frac{A_0+B_0}{(\xi_{0+}-\xi_{0-})+i\cot{\delta_{even}}} \\
          &   & \hspace{6em}  \mp \frac{A_0-B_0}{(\xi_{0+}-\xi_{0-})+i\cot{\delta_{odd}}},\nonumber\\
\label{eq:f0-T}
f_{0-}^\pm & = & \frac{A_0+B_0}{(\xi_{0+}-\xi_{0-})+i\cot{\delta_{even}}} \\
          &   & \hspace{6em}  \mp \frac{A_0-B_0}{(\xi_{0+}-\xi_{0-})+i\cot{\delta_{odd}}},\nonumber
\end{eqnarray}
\end{subequations}
where the 1D phase-shifts $\delta_{even}$ and $\delta_{odd}$ are defined by
\begin{subequations}
\begin{eqnarray}
\label{eq:deltag}
\cot{\delta_{even}} & \equiv & \left[(K\cot{\delta_0})d_U \right.\\
               &        & \hspace{4em} + \left. (C^2-d_U^2k_0^2)^{1/2}\right]\frac{k_0d_U}{2},\nonumber\\
\label{eq:deltau}
\cot{\delta_{odd}} & \equiv & \left[(K^3\cot{\delta_1})d_U^3 \right. \\
               &        & \hspace{4em} - \left. (C^2-d_U^2k_0^2)^{3/2}\right]\frac{1}{6k_0d_U}, \nonumber
\end{eqnarray}
\end{subequations}
with $C\equiv \sqrt{(q_1^2-q_0^2)d_U^2}$.

In solving Eq.(\ref{eq:Psi0border}), we still need to calculate $\xi_{0\pm}$. In this regard, note that the above
solutions for $f_{0+}^\pm$ and $f_{0-}^\pm$ make more apparent the fact that the poles of $f_{0+}^\pm$ and $f_{0-}^\pm$
for imaginary $k_0\equiv ik_{0B}$ are equal, something that was implicit already in Eqs.(\ref{eq:f0+}) and
(\ref{eq:f0-}). Such a pole is related to the bound state spectrum (see also discussion in Sec.V.E in~\cite{kim2005}),
in the sense that the first term on the rhs of Eq.(\ref{eq:Psi0border}) (the propagating part) becomes negligible
relative to the interacting terms in $e^{\pm k_{0B}|z|}$. If one wants then to retain the single site picture of a
bound-state-like exponentially decaying tail in the limit $a\rightarrow\infty$, it follows that $\xi_{0-}$ must
vanish in order to kill the diverging term $e^{+k_{0B}|z|}$ brought about by $f_{0-}^\pm$. Hence, by virtue of such
boundary conditions about bound states in the limit $a\rightarrow\infty$, we set $\xi_{0-}=0$ and $\xi_{0+}=1$, such
that Eq.(\ref{eq:Psi0border}) for $|z|\sim a/2$ becomes finally 
\begin{subequations}
\begin{equation}
\label{eq:Psi0end}
\Psi_0(\bm r)\approx A_0\,\psi_L(z)\,\varphi_0(\rho) + B_0\,\psi_R(z)\,\varphi_0(\rho),
\end{equation}
where $\psi_{L(R)}(z)$ describe a quasi-1D scattering of particles coming from the left (right) and are given by
\begin{eqnarray}
\label{eq:psiL}
\psi_L(z) & \equiv & \left\{ \begin{array}{ll}
                          e^{ik_0z} + A_r\,e^{-ik_0z}, & z\sim -a/2 \\
                          A_t\,e^{ik_0z}, & z\sim +a/2 
                        \end{array}
                     \right. \\
\label{eq:psiR}
\psi_R(z) & \equiv & \left\{ \begin{array}{ll}
                          A_t\,e^{-ik_0z}, & z\sim -a/2 \\
                          e^{-ik_0z} + A_r\,e^{ik_0z}, & z\sim +a/2
                        \end{array}
                     \right.
\end{eqnarray}
the quasi-1D scattering amplitudes $A_{t(r)}$ being defined by
\begin{eqnarray}
\label{eq:At}
A_t & \equiv & 1 - \frac{1}{1 + i\cot{\delta_{even}}} - \frac{1}{1 + i\cot{\delta_{odd}}}, \\
\label{eq:Ar}
A_r & \equiv &   - \frac{1}{1 + i\cot{\delta_{even}}} + \frac{1}{1 + i\cot{\delta_{odd}}}
\end{eqnarray}
and satisfying the conservation condition 
\begin{equation}
\label{eq:1Dcontinuity}
|A_t|^2 + |A_r|^2 = 1,
\end{equation}
\end{subequations}
valid for all real $\delta_{even}$ and $\delta_{odd}$, as can be verified explicitly. This constitutes the general
local solution at the borders of the site at the origin. The constants $A_0$ and $B_0$ are determined below from
Bloch's condition. 

\section{\label{sec:bands}Energy Bands under Confinement}
In order to obtain the solution $\Psi(\bm r)$ valid around all other sites, it is necessary to recall that it must
satisfy well known continuity conditions. In our case, due to the way this solution is being constructed, they should be
set at the site boundaries. For the site at the origin, for instance, one has at its right boundary at $z = +a/2$
\begin{subequations}
\begin{eqnarray}
\label{eq:continuity-a}
\Psi(x, y, a^-/2) & = & \Psi(x, y, a^+/2), \\
\label{eq:continuity-b}
\frac{\partial}{\partial z}\Psi(x, y, a^-/2) & = &
                                      \frac{\partial}{\partial z}\Psi(x, y, a^+/2),
\end{eqnarray}
\end{subequations}
where $a^-/2$ denotes the limit $z\rightarrow +a/2$ from the left and $a^+/2$ denotes this limit from the right. We now
use $\Psi_0(\bm r)$ in Bloch's equation Eq.(\ref{eq:bloch}) for $n_3=0$ ($\bm r$ around the site at the origin) and for
$n_3=+1$ ($\bm r$ around the first site to the right), thus
\begin{equation}
\label{eq:PsiPsi0}
\Psi(x,y,z) = \left\{ \begin{array}{ll}
                        \Psi_0(x,y,z), & |z| < a/2, \\
                        e^{ika}\Psi_0(x,y,z-a), & |z-a|< a/2.
                      \end{array}
              \right.
\end{equation}
Clearly, the first line should be used on the lhs of Eqs.(\ref{eq:continuity-a}) and (\ref{eq:continuity-b}) and the
second line on their rhs, so that
\begin{subequations}
\begin{eqnarray}
\label{eq:continuity-bloch-a}
\Psi_0(x, y, a/2) & = & e^{ika}\Psi_0(x, y, -a/2), \\
\label{eq:continuity-bloch-b}
\frac{\partial}{\partial z}\Psi_0(x,y,a/2) & = & 
                                      e^{ika}\frac{\partial}{\partial z}\Psi_0(x,y,-a/2).
\end{eqnarray}
\end{subequations}
For the next boundary at $z=+3a/2$, one uses $\Psi(\bm r)=e^{ika}\Psi_0(\bm r - a\bm e_z)$ for the left limit of
$z\rightarrow +3a/2$ (i.e. $\bm r$ around the site at $n_3=+1$) and $\Psi(\bm r)=e^{i2ka}\Psi_0(\bm r - 2a\bm e_z)$ for
the right limit (i.e. $\bm r$ around the site at $n_3=+2$); the continuity conditions then turn out to be the same, as it
should. In other words, Bloch's condition garantie the continuity along the whole lattice.

Using now Eq.(\ref{eq:Psi0end}), it can be seen that Eqs.(\ref{eq:continuity-bloch-a}) and (\ref{eq:continuity-bloch-b})
become an homogeneous system of equations for $A_0$ and $B_0$. This system will have a non-zero solution only if its
determinant vanishes, that is,
\begin{eqnarray}
\label{eq:determinant}
0 & = & \left(\psi_{L+}-e^{ika}\psi_{L-}\right) \left(\psi_{R+}'-e^{ika}\psi_{R-}'\right) \nonumber \\
  &   & \mbox{} -  \left(\psi_{L+}'-e^{ika}\psi_{L-}'\right) \left(\psi_{R+}-e^{ika}\psi_{R-}\right),
\end{eqnarray}
with $\psi_{L(R)\pm}\equiv \psi_{L(R)}(\pm a/2)$, $\psi_{L(R)\pm}'\equiv \psi_{L(R)}'(\pm a/2)$. Multiplying this
equation by $e^{-ika}$ and using Eqs.(\ref{eq:psiL}) and (\ref{eq:psiR}), one obtains after a tedious but
straightforward algebra
\begin{equation}
\label{eq:bands1}
\frac{A_t^2 - A_r^2}{2A_t}\,e^{ik_0a} + \frac{1}{2A_t}\,e^{-ik_0a} = \cos{(ka)}. 
\end{equation}
This equation can be written differently. From Eqs.(\ref{eq:At}) and (\ref{eq:Ar}), one can check that
$A_rA_t^\ast=i\alpha$ for real $\alpha$, i.e. this product is purely imaginary. Introducing the phase $\delta$ of $A_t$ 
\begin{equation}
\label{eq:delta}
A_t \equiv |A_t|\,e^{i\delta},
\end{equation}
it follows that $A_r=i\alpha e^{i\delta}/|A_t|$. Using $|A_rA_t^\ast| = |i\alpha|$ to extract $|A_t|$, one gets 
$A_r = i\alpha |A_r|\,e^{i\delta}/|\alpha|$. Taking this $A_r$ and Eq.(\ref{eq:delta}) into Eq.(\ref{eq:bands1}) and using
Eq.(\ref{eq:1Dcontinuity}) gives thus
\begin{subequations}
\begin{equation}
\label{eq:bands2}
\frac{\cos{(k_0a + \delta)}}{|A_t|} = \cos{(ka)},
\end{equation}
valid under the single channel constraint Eq.(\ref{eq:singlechannel}), which can be rewritten as
\begin{equation}
\label{eq:k0range}
0\leq k_0d_U \leq C.
\end{equation}
\end{subequations}
Eq.(\ref{eq:bands2}) is the desired equation that 
determines the energy band structure by giving $k_0$ or $E \equiv \hbar^2(q_0^2+k_0^2)/2m$ as a function of the
crystal momentum $k$. Not incidentally, this is nearly the same band structure equation as for the pure 1D case,
for which $V(r)$ is replaced by a pure 1D potential $V_{1D}(z)$ (see e.g. Eq.(8.76), Chap.8, problem 1, p.148
in~\cite{ashcroft1976} or~\cite{negretti2014}).

\begin{figure}[t]
\includegraphics[scale=0.68]{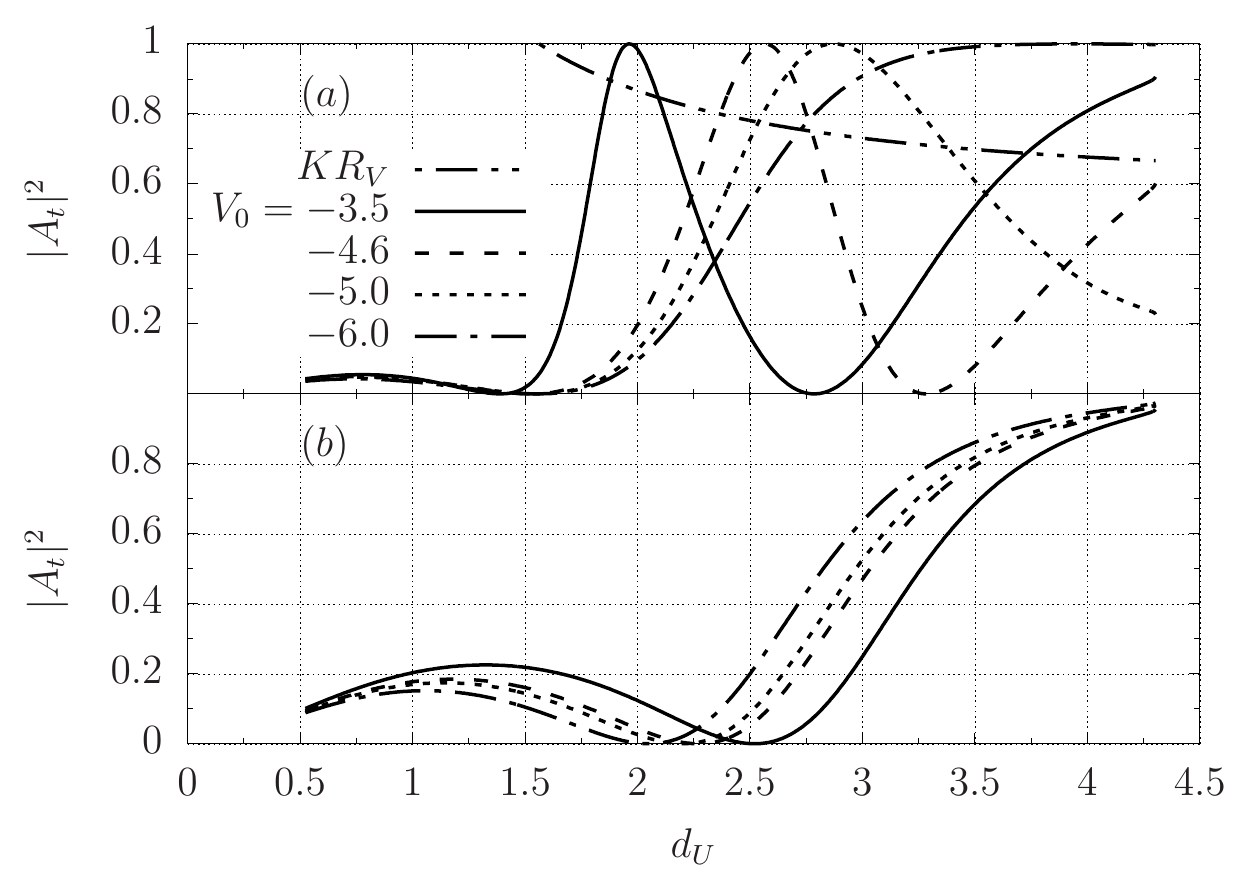}
\caption{\label{fig:At2dU} Transmission coefficient $|A_t|^2$ as a function of the confinement length scale $d_U$
  (in units of $R_V$) for several values of the potential depth $V_0$ (in units of $\hbar^2/mR_V^2$), with 
  $C \approx 2.58$ (square-well confinement $U$) and $k_0R_V = 0.60$. The range 
  $0.52 \approx 1/N_0 < d_U/R_V < C/k_0R_V \approx 4.3$ follows from $R_V < R_U$ and Eq.(\ref{eq:k0range}).
  (a) Both $s$- and $p$-waves are considered and one can see the CIRs for which $|A_t|=0$ and also the
  dual CIRs for which $|A_t| = 1$. The curve at the top shows how small $KR_V$, given by $K^2 = q_0^2 + k_0^2$, is
  from unity (low energy condition). (b) Only the $s$-wave contribution is considered. Note that then 
  $|A_t|\rightarrow 1$ only asymptotically.} 
\end{figure}
\begin{figure}[t]
\includegraphics[scale=0.7]{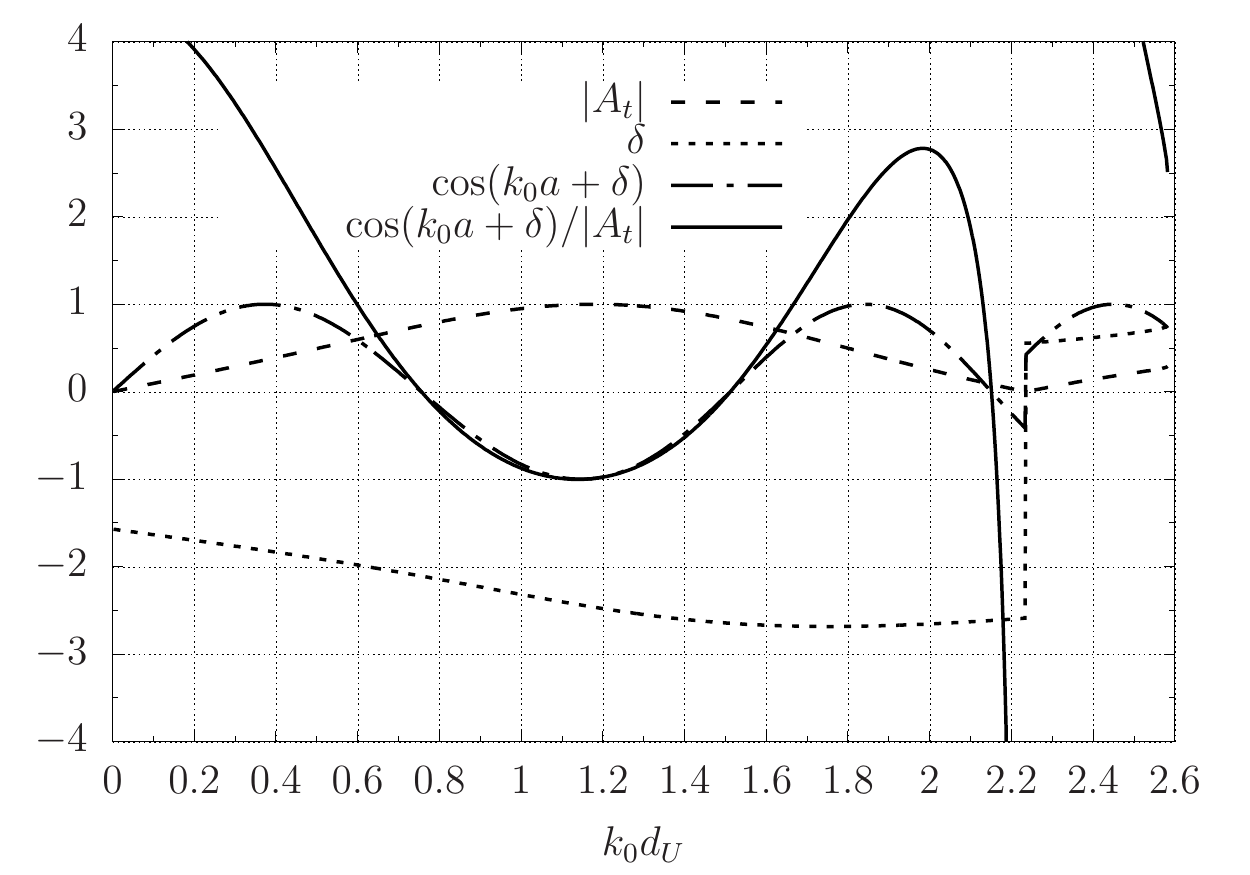}
\caption{\label{fig:lhs} An example of the elements of the lhs of Eq.(\ref{eq:bands2}) as a function of the
  longitudinal energy $k_0d_U$, whose range follows from Eq.(\ref{eq:k0range}), using 
  $C\approx 2.58$, $V_0 = -4.6\hbar^2/mR_V^2$, $a = 14.6R_V$ and $d_U = 3.00R_V$. Both $s$- and $p$-waves are
  included. One can see a zero gap for $k_0d_U$ between 1 and 1.2 when an extremum of $\cos{(k_0a + \delta)}$
  coincides with a dual CIR $|A_t| = 1$. The discontinuity in $\delta$ (set to the interval $-\pi$ to $\pi$) stems 
  from how $A_t$ approches and leaves the origin, for instance, from the third towards the first quadrant of the
  complex $A_t$ plane.}  
\end{figure}
The key difference, however, is that $A_t$ here is critically dependent on the confinement, more specifically, on
the parameter $d_U=R_U/N_0$ in the present case. The single channel constraint Eq.(\ref{eq:singlechannel}) implies
the low energy condition $K\sim q_0\sim 1/R_U \ll 1/R_V$, so that (see e.g. \cite{landau1980}, \S 132) 
\begin{subequations}
\begin{equation}
\label{eq:phaseshifts}
K\cot{\delta_0}\approx -1/a_s \hspace{1em} \mathrm{and} \hspace{1em} K^3\cot{\delta_1}\approx - 1/a_p^3,
\end{equation}
where $a_s$ and $a_p$ are the so-called (three-dimensional) $s$- and $p$-wave scattering lengths. Energy-dependent
corrections to Eq.(\ref{eq:phaseshifts}) may be accounted for but for simplicity we take $a_s$ and $a_p$ as
constants. As a concrete example, we may assume $V$ to be a spherical well of depth $V_0 < 0$ and radius $R_V$, so
that one can calculate 
\begin{eqnarray}
\label{eq:as}
a_s & = & \left( 1 -\frac{\tan{\xi}}{\xi}\right)R_V, \\
\label{eq:ap}
a_p^3 & = & \left( \frac{1}{3} - \frac{1 - \xi\cot{\xi}}{\xi^2}\right)R_V^3,
\end{eqnarray}
\end{subequations}
where $\xi\equiv (-2mR_V^2V_0/\hbar^2)^{1/2}$. For $|A_t|<1$ in Eq.(\ref{eq:bands2}), including the condition 
$|A_t| = 0$ called confinement induced resonance CIR~\cite{olshanii1998,dunjko2011,granger2004}, band gaps are
expected to appear, whereas {\em zero gaps} would require the dual CIR condition $|A_t|=1$, which is more suitable
to appear when both $s$- and $p$-wave contributions are taken into account~\cite{kim2006,kim2007}. The transmission
coefficient $|A_t|^2$ is shown in Fig.~\ref{fig:At2dU} for (a) both $s$- and $p$-wave contributions and (b) only
the $s$-wave contribution. An example of the influence of $A_t$ is shown in Fig.~\ref{fig:lhs}, illustrating a
typical behavior of the lhs of Eq.(\ref{eq:bands2}) for which a zero gap is expected to appear.

\begin{figure}[t]
\includegraphics[scale=0.68]{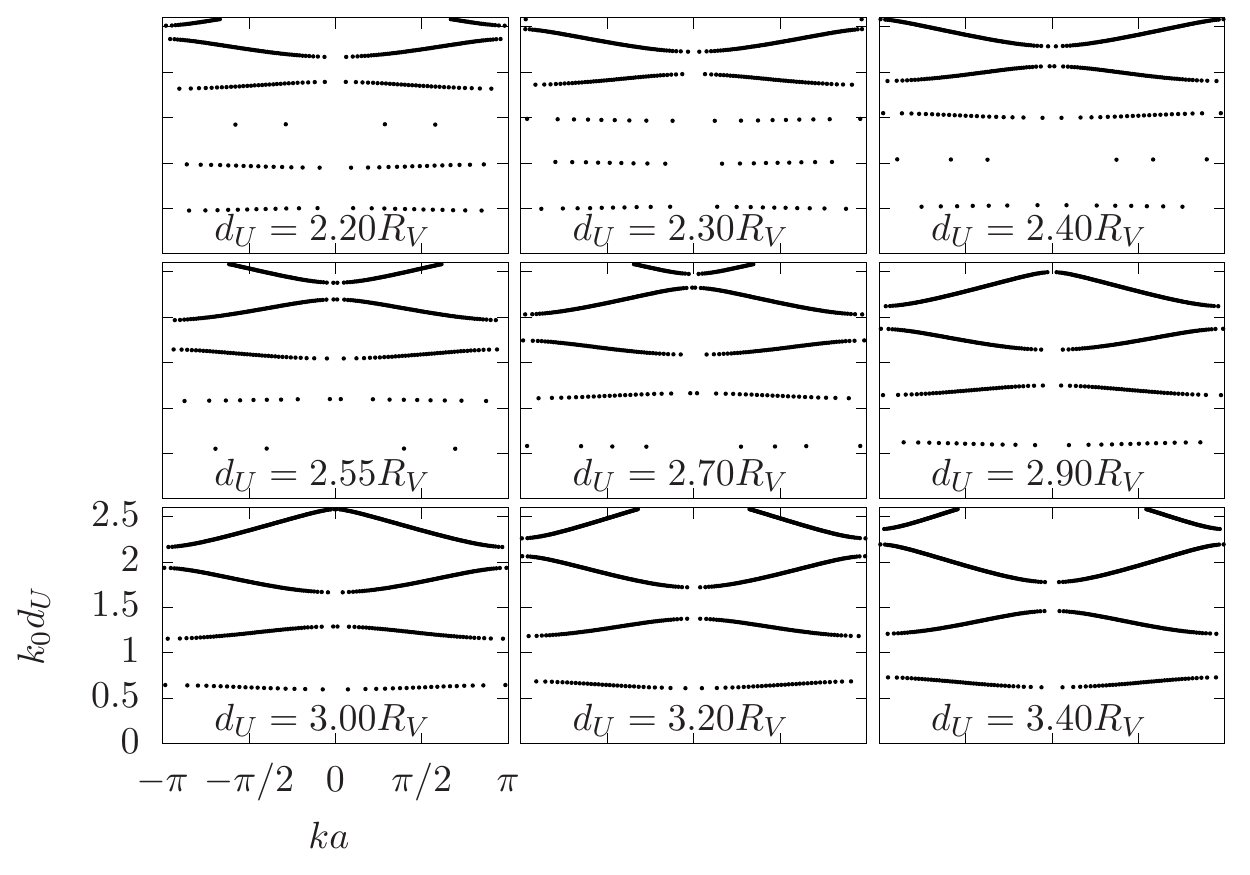}
\caption{\label{fig:bandsas} Band structure for various values of the confinement length scale $d_U$ when only
  the $s$-wave contribution is included. All graphs have the same axes. As expected, the bands do change under the
  influence of the confinement. However, closing a gap is hardly achieved.}
\end{figure}
\begin{figure}[t]
\includegraphics[scale=0.68]{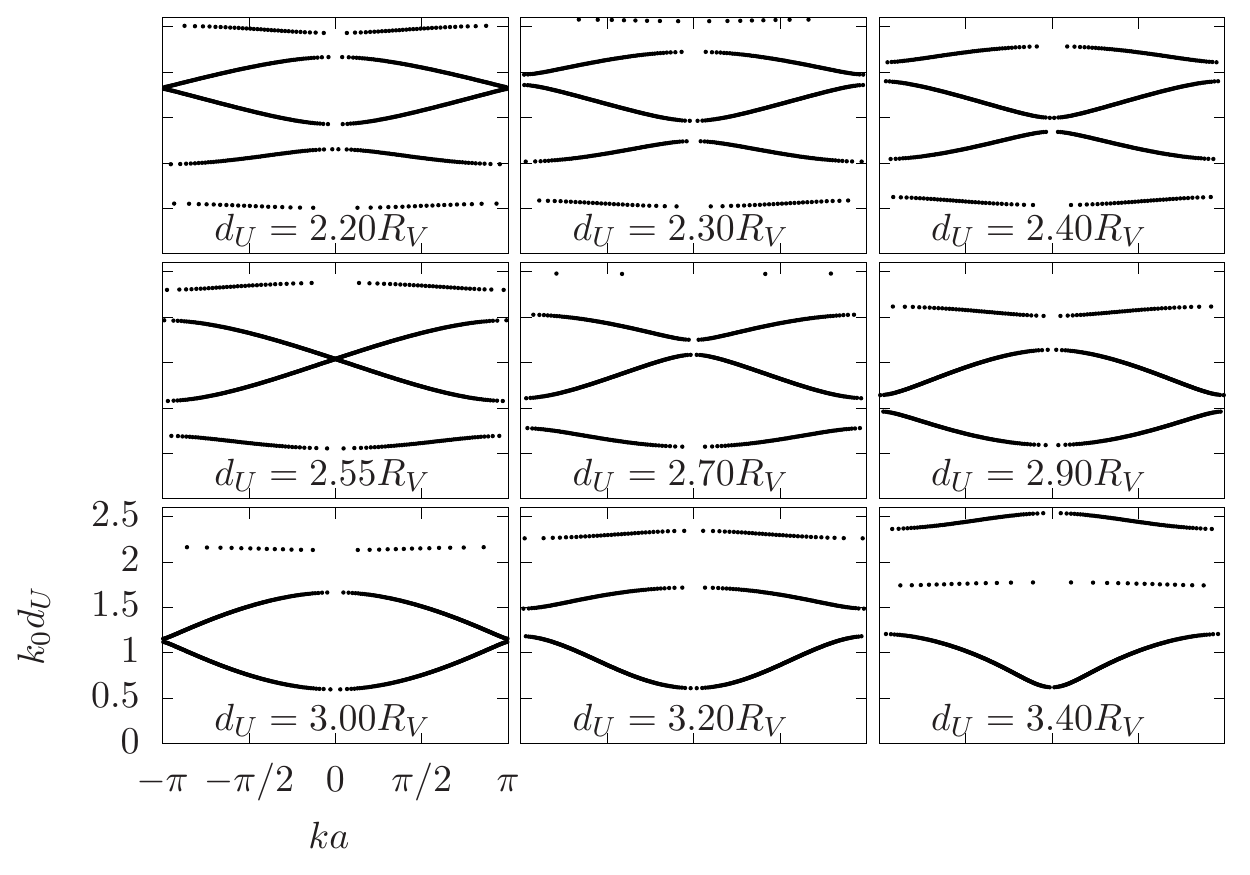}
\caption{\label{fig:bandsasap} The same as in Fig.~\ref{fig:bandsas}, but including both the $s$- and
  $p$-waves. The difference here is that, besides more noticeable changes, zero gaps or nearly so do appear, either
  at the center or at the edges of the quasi-momentum $k$-space (first column of graphs), due to the appearence of
  dual CIRs $|A_t|=1$.} 
\end{figure}
When solving Eq.(\ref{eq:bands2}), it is more convenient to vary $k_0$ and express $k$ as a function of $k_0$ and
then invert the relationship, which, however, results in few points for nearly flat bands. Using $C = 2.58$, 
$V_0 = -4.6\,\hbar^2/mR_V^2$ (such that both $a_s$ and $a_p$ are relatively large) and the lattice constant 
$a = 14.6\,R_V$, one can then obtain the bands for various values of $d_U$. This is done in Fig.~\ref{fig:bandsas},
where the $p$-wave contribution is suppressed. In Fig.~\ref{fig:bandsasap}, on the other hand, it is kept for the
sake of comparisson and a pratically zero gap can be observed, in addition to more pronounced changes to the
bandwidths. Although beyond the scope of the present work, it may be noted that the first band shows a significant
change as a function of $d_U$, which could be exploited for driving some parameters in a quasi-1D Bose-Hubbard
Hamiltonian approximation as a function of the physical 3D parameters.

\section{\label{sec:discussion}Discussion}
The possibility of openning and completely closing a band gap continuously as shown in Fig.~\ref{fig:bandsasap}
raises the prospect of dynamically driving a material between a (gap) insulator-like and a metal-like behavior if
all other conditions are met such as the proper distribution of particles along the bands.

As trial systems, one may consider ionized impurities regularly placed along the axis of a semiconductor quantum
wire surrounded by hard walls with different radii (see e.g. discussion at the end of Ref.~\cite{kim2006} and
references therein) or atom-ion systems as already suggested in~\cite{negretti2014} (see also~\cite{melezhik2016})
for which trapped ions would form the lattice and the atom would move in an atom waveguide obtained, e.g., with an
optical potential. An all optical system could also be tried.

Indeed, consider in more detail such optical potentials, which have the advantage of allowing to dynamically change 
the confinement in contrast to hard walls. The effective potential felt by a cold neutral atom, after time
averaging over the fast optical oscillations, is proportional to the intensity of the laser 
field~\cite{grimm2000,lewenstein2012}. Three pairs of counter-propagating laser beams with frequency $\omega_L$ and
the same electric field amplitude $E_L$, with each pair along one of the three orthogonal axes and with suitable
linear polarizations, can be made to yield the net electric field
\begin{equation*}  
2E_L [\sin{(k_Lx)}{\bm e_2} + \sin{(k_Ly)}{\bm e_3} + \sin{(k_Lz)}{\bm e_1}]\cos{(\omega_L t)}
\end{equation*}
which, after time averaging over a period $T\approx 2\pi/\omega_L$, generates a 3D lattice potential
\begin{equation}
\label{eq:vlatt}
\tilde{V}_{latt}({\bm r}) = V_L\,[\,\sin^2{(k_Lx)} + \sin^2{(k_Ly)} + \sin^2{(k_Lz)}\,]
\end{equation}
where $V_L$ is proportional to $2E_L^2$. Adding then two other pairs of beams with frequency $\omega_C$ and
amplitude $E_C$ such as to provide the field (which alone would generate a waveguide-like potential)
\begin{equation*}
2E_C [\sin{(k_Cx)}{\bm e_2} + \sin{(k_Cy)}{\bm e_3}]\cos{(\omega_C t)},
\end{equation*} 
the total electric field of these five pairs of beams yields then the total optical potential
\begin{eqnarray}
\lefteqn{V_{T}({\bm r}) = \tilde{V}_{latt}({\bm r})} \\ 
               &  & \mbox{} + V_C\sin^2{(k_Cx)} + 2\sqrt{V_LV_C}\sin{(k_Lx)}\sin{(k_Cx)} \nonumber\\
               &  & \mbox{} + V_C\sin^2{(k_Cy)} + 2\sqrt{V_LV_C}\sin{(k_Ly)}\sin{(k_Cy)} \nonumber
\end{eqnarray}
where $V_C$ is proportional to $2E_C^2$ and one assumes $\omega_C\approx \omega_L$, namely 
$|\omega_L - \omega_C| \ll \omega_L$. Close to the $z$-axis, one obtains approximately
\begin{equation}
V_T({\bm r}) \approx \tilde{V}_{latt}({\bm r}) + \tilde{U}(\rho)
\end{equation}
where 
\begin{equation}
\tilde{U}(\rho) \equiv \left(V_Ck_C^2 + 2\sqrt{V_LV_C}k_Lk_C\right)\rho^2
\end{equation}
would be the confining potential that could be tuned by varying the intensity $V_C$ and the wavelength 
$\lambda_C = 2\pi/k_C$, provided the atoms could be carefully loaded close to the $z$-axis.

It must be noted that some assumptions made here may not be assured, particularly Eq.(\ref{eq:approxRvRua})
(although stretching the lattice spacing $a$ may be attempted~\cite{fallani2005}) or the condition of
simultaneously large contributions of $s$- and $p$-waves. However, if not all aspects of the present discussion, at
least some of them, such as the continuous qualitative change the confinement may impose on the band structure, may
then be illustrated.

\begin{acknowledgments}
The author gratefully acknowledges fruitful discussions with Paulo A. Nussenzweig, Peter Schmelcher and Vladimir 
S. Melezhik and wish to thank Peter Schmelcher for reading an early version of the manuscript and for suggesting
many improvements and references, particularly Ref.~\cite{negretti2014}. Financial support by the Universidade
Federal de S\~ao Paulo is gratefully acknowledged.
\end{acknowledgments}

\bibliography{kim2020a}

\end{document}